\documentclass[pre,superscriptaddress,twocolumn,floatfix,longbibliography]{revtex4-1}

\usepackage{graphics}
\usepackage{graphicx}
\usepackage{amsmath}
\usepackage{amssymb,xcolor}
\usepackage{textcomp}
\usepackage{siunitx}
\usepackage{xcolor}
\usepackage{ulem}

\newcommand{\textin}[1]{\mbox{\scriptsize{#1}}}

\newcommand{\revFJGR}[1]{\textcolor{black}{#1}}

\definecolor{grisclair}{rgb}{0.6,0.6,0.6}

\newcommand{\beq}{\begin{equation}}
\newcommand{\ee}{\end{equation}}

\begin{document}

\title{Impulsive Hydrodynamic Exfoliation into Monolayer Graphene and Nanofragments by Transonic Flow Focusing}
\author{A. Ponce-Torres}
\address{Depto.\ de Ingenier\'{\i}a Mec\'anica, Energ\'etica y de los Materiales and\\ 
Instituto de Computaci\'on Cient\'{\i}fica Avanzada (ICCAEx),\\
Universidad de Extremadura, E-06006 Badajoz, Spain}

\author{A. Rubio-Gonz\'alez}
\address{Depto.\ de Ingenier\'{\i}a Mec\'anica, Energ\'etica y de los Materiales and\\ 
Instituto de Computaci\'on Cient\'{\i}fica Avanzada (ICCAEx),\\
Universidad de Extremadura, E-06006 Badajoz, Spain}

\author{J. M. Montanero}
\address{Depto.\ de Ingenier\'{\i}a Mec\'anica, Energ\'etica y de los Materiales and\\ 
Instituto de Computaci\'on Cient\'{\i}fica Avanzada (ICCAEx),\\
Universidad de Extremadura, E-06006 Badajoz, Spain}

\author{M. A. Herrada}
\address{E.T.S.I., Depto.\ de Ingenier\'{\i}a Aeroespacial y Mec\'anica de Fluidos, Universidad de Sevilla, Camino de los Descubrimientos s/n 41092, Spain}

\author{F. J. Galindo-Rosales}
\email{Correspondence and requests for materials should be addressed to galindo@fe.up.pt}
\thanks{A. Ponce-Torres and F.J. Galindo-Rosales contributed equally to this work.}
\address{CEFT, Departamento de Engenharia Qu\'{\i}mica e Biológica, Facultade de Engenharia da Universidade do Porto,\\
Rua Dr. Roberto Frias, 4200-465 Porto, Portugal}

\begin{abstract}
We propose using Transonic Flow Focusing (TFF) to produce 2D and 0D nanomaterials. This technique focuses liquid suspensions into high-speed micrometer-scale jets, combining extremely high shear and elongational stresses in a confined, contact-free zone. \revFJGR{For the Graphene Nanoplatelets suspensions and TFF operating conditions investigated here, the process promoted exfoliation without added surfactants or oxidative chemistry}. Both graphene monolayers \revFJGR{flakes} ($\sim 300-400$ nm in lateral size) and monolayer graphene \revFJGR{nanofragments with lateral sizes compatible with quantum dots} ($\sim 10-15$ nm) were obtained in a single \revFJGR{TFF} step using isopropanol and pure water. Our theoretical analysis reveals that\revFJGR{, during microsecond residence times at the meniscus-jet transition,} shear and extensional stresses of the order of $10^6$ s$^{-1}$ \revFJGR{act on the suspended particles,} yield\revFJGR{ing} viscous power densities of the order of $10^{10}$~$\mathrm{W/m^{3}}$.  High-resolution transmission electron microscopy and atomic force microscopy show that the monolayer fraction exceeded 99\% for isopropanol and 92.9\% for water. \revFJGR{These results suggest that TFF can combine solvent versatility with a high monolayer fraction in a purely mechanical top-down process.}
\end{abstract}
\maketitle

\section{Introduction}

Two-dimensional (2D) materials, especially graphene, have redefined the frontiers of materials science. Graphene is a single layer of carbon atoms arranged in a hexagonal lattice, just 0.33 nm thick, but with \revFJGR{exceptional} properties: quantum confinement \citep{ZTSK05,NMMFKZJSG06,NJZMSZMBKG07,NGMJKGDF05}, ultrahigh electrical \citep{MGMBJPBNWTG11} and thermal conductivity \citep{BGBCTML08}, optical transparency \citep{NBG08}, and exceptional mechanical strength \citep{LWKH08}. These properties underpin applications in nanoelectronics \citep{AKLRR16}, quantum devices \citep{LRSSZ15}, biosensors \citep{P11}, flexible displays \citep{KCLKH15}, composites \citep{GB20}, or smart textiles \citep{CKLCSC25}. However, the most \revFJGR{prominent} electronic and mechanical phenomena are strictly layer-dependent, typically reaching their theoretical limits only in the pristine monolayer regime \citep{PKJPM22,P24}. Therefore, the ability to isolate single-layer graphene is essential for \revFJGR{enabling} both fundamental research and real-world applications \citep{SBMA24}. 

Graphene Quantum Dots (GQDs) \citep{GHKSDAKA22} are nanofragments of monolayer or few-layer graphene, typically below 30 nm in lateral size, that exhibit size-dependent photoluminescence due to quantum confinement. They offer unique prospects for quantum optics, bioimaging, drug delivery, and optoelectronics \citep{GHKSDAKA22}.

\revFJGR{Producing high-quality graphene monolayers~\citep{SBMA24} and graphene nanofragments~\citep{GHKSDAKA22} in a clean, scalable, and cost-effective manner remains challenging.} Most methods are either chemically invasive \citep{AKLRR16,GHKSDAKA22} (utilising oxidation–reduction cycles or acids) or energy-intensive \citep{LRSSZ15,PKJPM22,GHKSDAKA22}. Additionally, they are often incompatible with downstream applications \citep{P24,GHKSDAKA22,XCXLC18}. Chemical Vapour Deposition (CVD) methods grow graphene from gaseous carbon sources on catalytic metal substrates at high temperatures \citep{AKLRR16,PKJPM22,CYWOKKL22,JZZSLL21}. These techniques can produce large-area monolayers with excellent structural quality and are the method of choice for high-performance electronic applications \citep{CYWOKKL22}. However, CVD is energy-intensive \citep{JZZSLL21,SRCI18}, slow \citep{JZZSLL21,SRCI18}, and costly \citep{JZZSLL21,SRCI18,BBRC15}, and requires delicate transfer steps to remove the graphene from the substrate \citep{P24,CYWOKKL22,JZZSLL21}. These steps often introduce cracks, wrinkles, or contamination, which limit device integration and severely hinder scalability \citep{P24,CYWOKKL22,JZZSLL21}. Additionally, it requires large quantities of hydrogen, raising safety concerns \citep{SRCI18}. 

Liquid-phase exfoliation (LPE) methods utilise hydrodynamic forces to disperse a suspension of layered materials, thereby overcoming van der Waals forces that hold layers together to delaminate flakes into monolayers. It is one of the most scalable and versatile approaches and can be applied to a wide range of 2D materials \citep{XCXLC18,CYWOKKL22}. However, these methods are energy-intensive \citep{XCXLC18,KU24} and slow \citep{XCXLC18,KOLDC10}, and the monolayer fraction is typically low \citep{SPUMP18,XCXLC18,WYH23}. The inherent stochastic nature of energy dissipation in bulk LPE (e.g., during sonication or high-shear mixing) results in a broad distribution of flake thicknesses, with the monolayer fraction rarely exceeding 20\% without intensive, yield-limiting post-processing such as liquid cascade centrifugation (LCC). In fact, the resulting dispersions contain flakes with a broad distribution of sizes and thicknesses. Moreover, these methods often work under aggressive conditions (high pressure, impacts, local overheating, shock waves, and grinding), which introduce significant structural defects \citep{XCXLC18,CYWOKKL22}. Extensional flows in convergent–divergent nozzles have been proposed for this purpose because they are more effective than shear flows \citep{XDLCCHCZL22}. These microfluidic systems have not evolved beyond their initial application to high-viscosity polymer-based nanocomposites \citep{XDLCCHCZL22,AMM21}, and their further improvement is strongly constrained by nozzle fabrication.  

The situation is even more critical at the 0D limit (GQDs). Top-down methods, such as oxidative cutting, strong sonication, or laser ablation, introduce structural defects and broad size distributions. They often rely on harsh experimental conditions, including strong acids, high voltages, or prolonged reaction times, which are incompatible with large-scale production \citep{GHKSDAKA22,TL24,TMKMM21}. Bottom-up approaches, which assemble GQDs from molecular precursors, typically require several hours of synthesis at high temperatures and pressures and generally have low throughput \citep{GHKSDAKA22,TL24,TMKMM21}. In addition, they often produce substantial by-products, making purification of the as-synthesized GQDs a major challenge \citep{GHKSDAKA22}. 

In this paper, we propose using Transonic Flow Focusing (TFF) to produce 2D and 0D nanomaterials. In contrast to the \revFJGR{broadly distributed} stochastic energy distribution in bulk LPE, TFF provides a \revFJGR{spatially localized and reproducible hydrodynamic} environment, enabling a wall-free, mechanical \revFJGR{top-down} platform for the scalable production of 2D and 0D nanomaterials.

\section{Transonic Flow Focusing}

\begin{figure}
\begin{center}
\resizebox{0.4\textwidth}{!}{\includegraphics{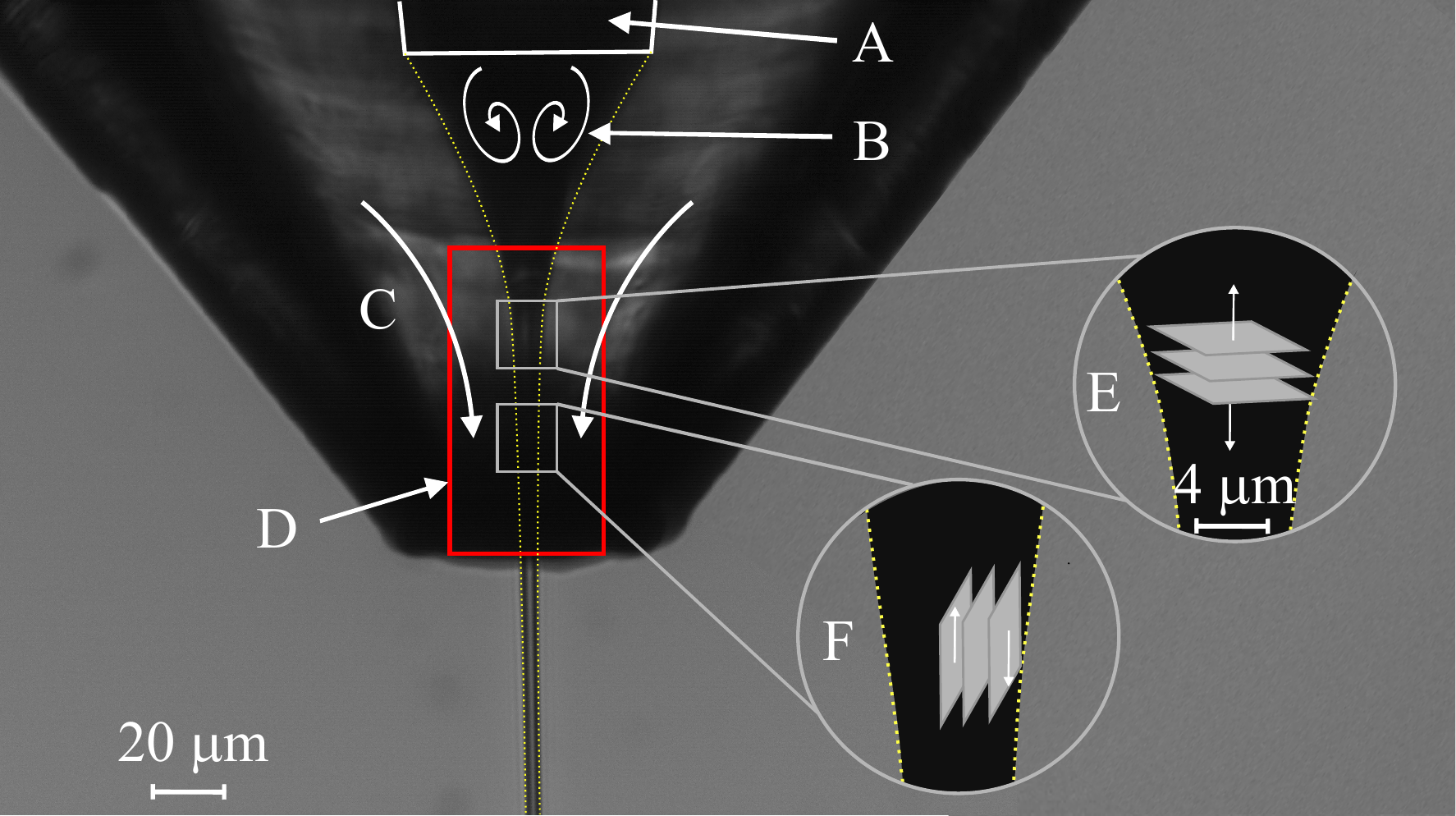}}
\end{center}
\vspace{0cm}
\caption{Transonic Flow Focusing. The arrows show the inner capillary (A), the recirculation cells in the liquid meniscus (B), the transonic air stream (C), and the cone-jet transition (D). Extensional (E) and shear (F) stresses contribute to the exfoliation of graphene layers.}
\label{sketch}
\end{figure}

In TFF \cite{G98a,Cetal11} (Fig.\ \ref{sketch}), a liquid stream is accelerated and thinned by a coflowing air transonic current through a submillimeter converging nozzle. A continuous micrometer jet tapers from the tip of a liquid meniscus. Inside the meniscus, two symmetric recirculation cells create a near-stagnant region in front of the jet. The thin boundary layer between the fast-moving air–liquid interface and the recirculation cells serves as a virtual channel where exfoliation initiates. However, the core of this deterministic mechanism lies in the meniscus–jet transition, where the liquid accelerates to speeds of up to 50 m/s over a few microns, creating localized, extreme peaks in shear and strain rates. This unique combination of orthogonal stresses provides a powerful \revFJGR{both extensional and shear contributions, increasing the probability of activating exfoliation for different particle orientations. }
\revFJGR{At the low flow rate used here, a large fraction of the liquid is expected to pass through the high-gradient meniscus–jet region}, ensuring high efficiency and consistent output. Crucially, no solid walls are involved during exfoliation, avoiding defects typically introduced by abrasion, friction, or mechanical impact. Moreover, no surfactants, additives, or post-processing steps are needed.

In our experiments, the pre-expanded graphene nanoplatelet (GNP) suspension (see Sec.\ \ref{mat:expmet}) was injected through the inner capillary of inner diameter $D_c=75$ $\mu$m, which was coaxially located within a converging nozzle. The capillary edge is located at a distance $H=180$ $\mu$m from the nozzle orifice, whose diameter is $D=75$ $\mu$m. The liquid was injected at a constant flow rate of $Q_l=1.5$ ml/h. The focusing air stream was established by setting an upstream stagnation pressure $p_0=2.5$ bar and temperature $T_0=293$ K. The ejector was mounted onto the cap of a discharge glass cell at atmospheric pressure. The suspension flow rate was maintained close to the minimum value for stable flow to maximize energy focusing in the meniscus tip \citep{M24}. Under these conditions, the resulting microjet had a diameter $d_j\simeq 3.8$ $\mu$m and a velocity $v_j\simeq 36$ m/s, remaining stable for over one hour of continuous operation. Details of the experimental method can be found in Sec.\ \ref{mat:expmet}.

\section{Hydrodynamic Mechanism} 

To elucidate the physical origin of the \revFJGR{high} exfoliation efficiency observed in TFF, we analyzed the flow field from both a scaling analysis and numerical simulations.

\subsection{Scaling Analysis}
\label{sec:scaling}

The average strain rate in the liquid meniscus can be estimated as $\langle \dot{\varepsilon}\rangle\sim v_j/D\simeq 0.48\times 10^6$ s$^{-1}$, where $D$ is an estimate of the meniscus length. Assuming uniaxial extensional flow in the meniscus tip, the strain rate in this region can be estimated as $\dot{\varepsilon}=-2Q_l/[\pi r(z)^3] dr/dz$. Using simulation the simulation values $r(z)\simeq 8.5$ $\mu$m and $dr/dz\simeq -15$, we find that $\dot{\varepsilon}\sim 7\times 10^6$ s$^{-1}$. These results confirm that TFF operates in a regime in which extensional stresses are orders of magnitude higher than those in conventional LPE.

The transonic nature of the gas stream plays a critical role. At the nozzle orifice, the air current reaches the speed of sound $v_{\textin{air}}^*=313$ m/s. The pressure, temperature, and density corresponding to the sonic flow are $p^*=1.32$ bar, $T^*=244$ K, and $\rho^*=1.89$ kg/m$^3$. Using Sutherland's law to estimate the air viscosity $\mu_{\textin{air}}(T)$, the Reynolds number in the air stream at the nozzle orifice is estimated as $\text{Re}_x=x\rho^* v_{\textin{air}}^*/\mu_{\textin{air}}(T^*)\simeq 2820$, where we have considered $x\simeq D$. The air boundary layer can be estimated from Blasius' formula $\delta_{\textin{air}}^*=4.92\, D/\text{Re}_x^{1/2}\simeq 7$ $\mu$m. The shear viscous stress at the free surface is $\tau\sim \mu_{\textin{air}}v_{\textin{air}}^*/\delta_{\textin{air}}^*\simeq 707$ Pa. The stress balance at the interface yields $\tau=\mu_l \dot{\gamma}$, which leads to a shear rate $\dot{\gamma}\sim 10^6$ s$^{-1}$ on the liquid side of the interface \citep{V26}. Shear rates of this order of magnitude persist downstream until the jet velocity profile is fully developed. Consider, for instance, a velocity variation $\delta v_j$ of the order of 1 m/s in the radial direction. This leads to shear rates $\dot{\gamma}\sim \delta v_j/(d_j/2)\simeq 10^6-10^7$ s$^{-1}$ in the jet. This range of values is \revFJGR{consistent with the strong morphological transformation observed after TFF processing of the pre-expanded GNPs, as discussed in Sec.~\ref{IIIC}.}

It is also useful to estimate the viscous energy available in the flow and compare it with the exfoliation energy. Shear and extensional stresses at the meniscus-jet transition yield extremely high viscous power densities. This hydrodynamic energy is transmitted to the suspended pre-expanded particle over microsecond residence times. 

The cumulative viscous work per unit volume is $w_v=\int_{0}^{t} \boldsymbol\tau:\nabla{\bf v}\, dt'$, where $\mathbf{\tau}$ is the viscous stress tensor and $t'$ is the time over which the viscous forces act on the fluid particle. The scalar $\boldsymbol\tau:\nabla{\bf v}$ provides a measure of the mechanical \revFJGR{power} density \revFJGR{dissipated} in the flow\revFJGR{. Although it does not account for the coupling between the flow and platelets orientation, it is useful for identifying the regions where mechanical energy is most intensely supplied to the GNPs}. The work $w_v$ can be estimated as $w_v\sim \mu (\dot{\varepsilon}^2+\dot{\gamma}^2)t_{\textin{res}}$, where $\dot{\varepsilon}$ and $\dot{\gamma}$ are characteristic values of the strain and shear rates in the meniscus-jet transition. In addition, $t_{\textin{res}}$ is the residence time of the GNP stacks within the high-gradient region of the meniscus. Assuming $\dot{\varepsilon}\sim 10^6-10^7$, $\dot{\gamma}\sim 10^6-10^7$, and $t_{\textin{res}}\sim 10$ $\mu$s, we obtain $w_v\sim 10^5~\mathrm{J m^{-3}}$. \revFJGR{This quantity should therefore be interpreted as an upper-bound estimate of the viscous energy density that can be supplied by the surrounding flow during the residence time in the high-gradient region, rather than as the energy effectively transferred to a specific GNP stack.}

\revFJGR{A comparison with the interfacial energy required for complete layer separation is provided in the Supplementary Material. This comparison highlights that a volume-averaged energy balance is not a predictive exfoliation criterion, because the actual work transferred to a GNP stack depends on particle orientation, trajectory, edge delamination, pre-existing defects, and particle--flow interactions \citep{Gravelle2020JCP, Kamal2020}.}

\subsection{Numerical simulation}

To gain deeper insight into the hydrodynamic mechanism underlying monolayer exfoliation, we conducted numerical simulations of the flow in TFF with water. The governing equations and the numerical method used to integrate them are described in Sec.\ \ref{sec:MatMetNum}.

\begin{figure}
\begin{center}
\resizebox{0.42\textwidth}{!}{\includegraphics{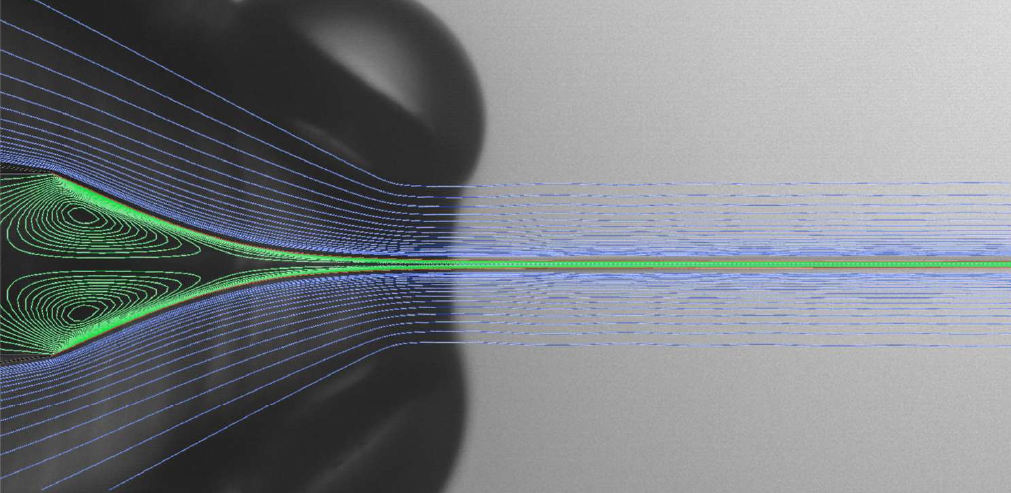}}
\resizebox{0.5\textwidth}{!}{\includegraphics{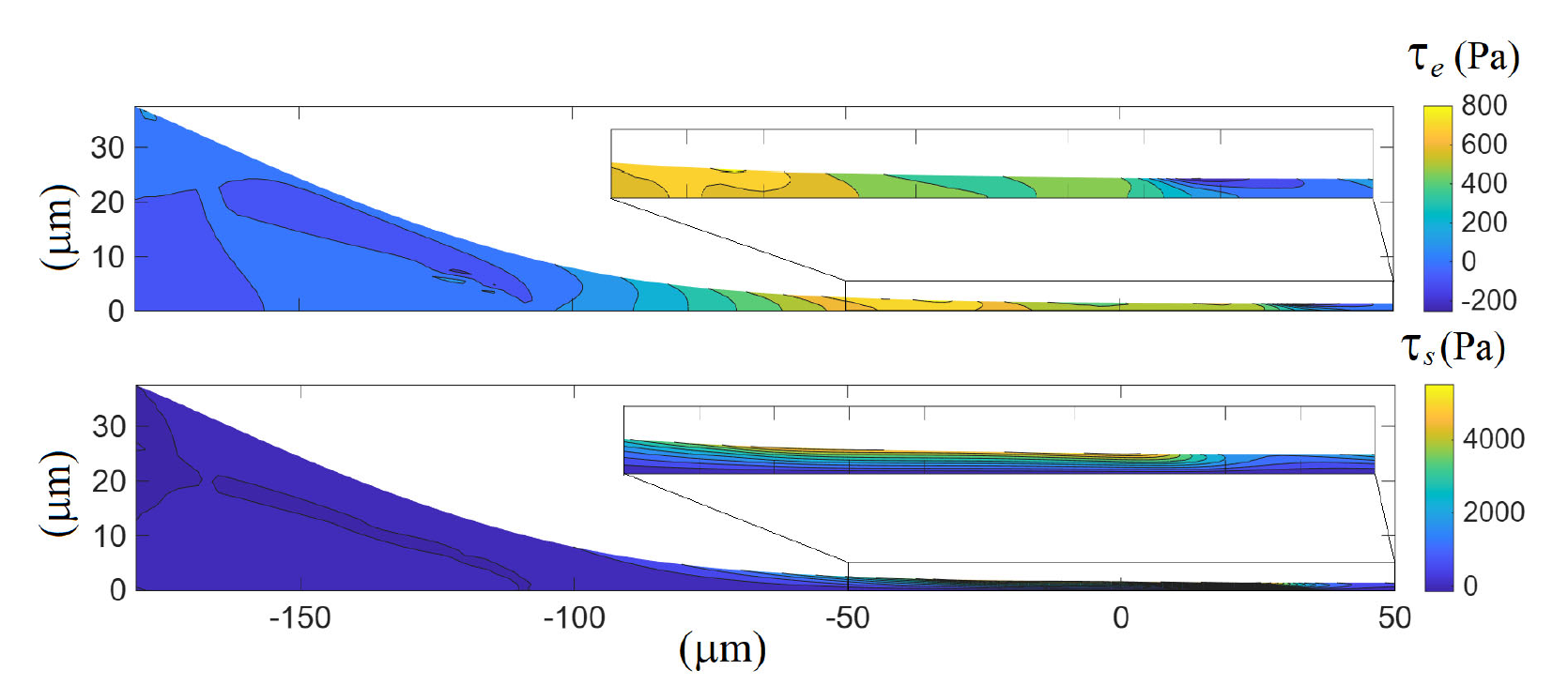}}
\end{center}
\vspace{0cm}
\caption{Streamlines obtained from the numerical simulation plotted on an experimental image. Contour plot of the extensional $\tau_e(r,z)$ and shear $\tau_s(r,z)$ stress obtained from the numerical simulation. The origin of the horizontal axis corresponds to the discharge orifice location. The numerical simulation was conducted for water and $Q_l=1.5$ ml/h.}
\label{stress}
\end{figure}

The remarkable agreement between the experimental and numerical liquid shapes suggests that the GNPs do not significantly alter the flow. The streamlines plotted in Fig.\ \ref{stress} show the strong recirculation pattern arising inside the liquid meniscus for the liquid flow rate considered in our experiments. This indicates that the flow rate is close to the minimum required for stable jetting. As anticipated, the thin boundary layer between the fast-moving air–liquid interface and the recirculation cells serves as a virtual channel through which the graphene flakes are ejected. The streamline topology indicates that particles do not follow simple axial trajectories, but instead experience complex paths involving recirculation zones and strong velocity gradients before entering the jet. The insets in Fig.\ 2b show that the fluid particle undergoes extensional stresses $\tau_e\sim 10^2-10^3$ Pa and shear stresses $\tau_s\sim 10^3-10^4$ Pa across a jet region with a length $\ell_j\sim 100$ $\mu$m. As anticipated above, the fluid particle residence time $t_{\textin{res}}$ within this region is of the order of $10$ $\mu$s.

Now we focus on the viscous stress responsible for exfoliation. Consider the unit vector ${\bf \hat{v}}={\bf v}/v$ parallel to the axisymmetric velocity field ${\bf v}(r,z)=v_r(r.z)\, {\bf e_r}+v_z(r,z)\, {\bf e_z}$. Let ${\bf n}=n_r {\bf e_r}+n_{\theta}{\bf e_{\theta}}+n_z{\bf e_z}$ be the unit vector perpendicular to one of the two faces of a thin parallepipedal flake. For simplicity, we will assume that $n_{\theta}=0$; i.e., the flake is not rotated around the flow symmetry axis. The unit vectors ${\bf n_{\parallel}}$ and ${\bf n_{\perp}}$ denote the directions parallel and perpendicular to the liquid velocity, respectively; i.e., ${\bf n_{\parallel}}\cdot {\bf \hat{v}}=1$ and ${\bf n_{\perp}}\cdot {\bf \hat{v}}=0$. The exfoliation mechanism is governed not only by the magnitude of the viscous stress tensor ($\boldsymbol{\tau}$), but critically by its projection onto the flake orientation. To capture this effect we will consider the two limiting flake configurations. When the flake normal is aligned with the flow direction (${\bf n}={\bf n_{\parallel}}$, see E in Fig.\ \ref{sketch}), the dominant contribution is extensional, $\tau_e={\bf n_{\parallel}}\cdot\boldsymbol{\tau}\cdot {\bf n_{\parallel}}$, promoting normal separation of stacked layers. In contrast, when the flake normal is perpendicular to the flow (${\bf n}={\bf n_{\perp}}$, see F in Fig.\ \ref{sketch}), the relevant contribution is shear, $\tau_s={\bf n_{\parallel}}\cdot\boldsymbol{\tau}\cdot {\bf n_{\perp}}$, which promotes interlayer sliding. 

Figure \ref{stress} shows the spatial distribution of the extensional and shear stresses. The computed values are fully consistent with the strain rate $\dot{\varepsilon}\sim \tau_e/\mu$ and shear rate $\dot{\gamma}\sim \tau_s/\mu$ ($\mu=1$ mPa$\cdot$s is the liquid viscosity) estimated from the scaling analysis. Strain rates of the order of $10^6$ s$^{-1}$ ($\tau_e\sim 10^3$ Pa) are produced across the meniscus tip, in front of the discharge orifice. The highest values of the shear rate are obtained near the jet surface in the discharge orifice region due to the action of the transonic air stream. The value of $\tau_s$ lies in the interval $\dot{\gamma}\sim 10^6-10^7$ s$^{-1}$ ($\tau_s\sim 10^3-10^4$ Pa). The shear stress $\tau_s$ considerably decreases for distances from the discharge orifice larger than $\sim 50$ $\mu$m due to viscous diffusion of the axial momentum across the jet section.  

\begin{figure}
\begin{center}
\resizebox{0.45\textwidth}{!}{\includegraphics{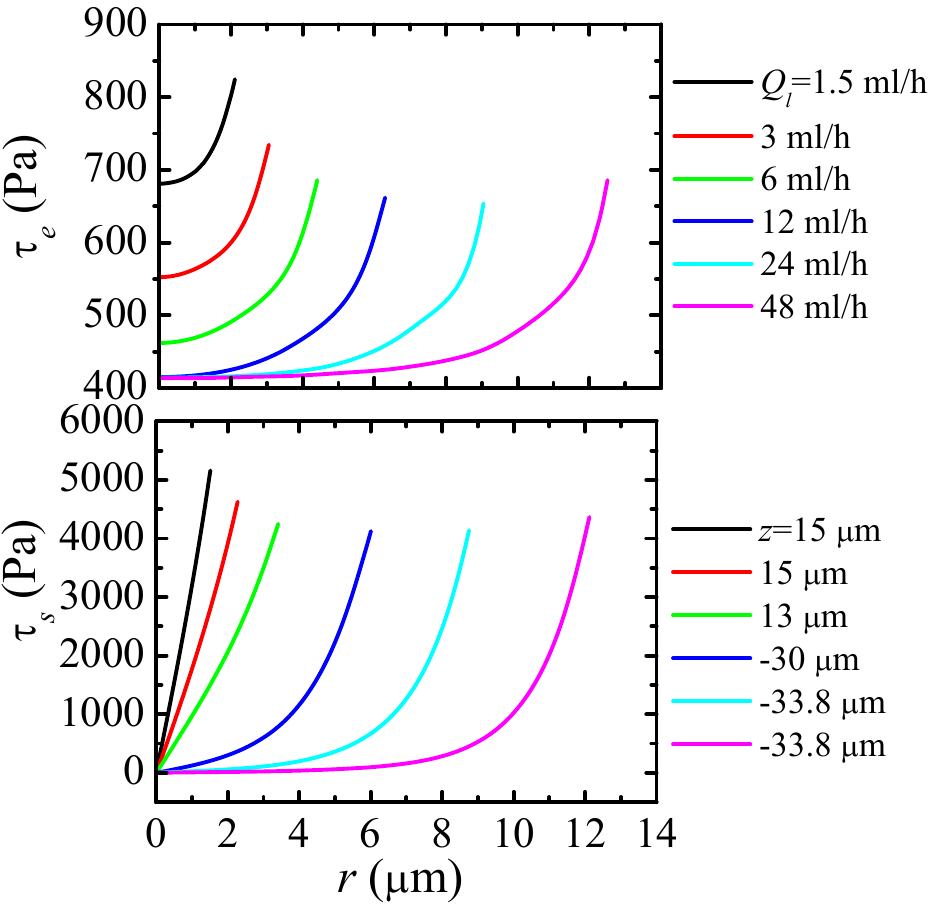}}
\end{center}
\vspace{0cm}
\caption{Extensional $\tau_e$ and shear $\tau_s$ stress as a function of the radial coordinate $r$ evaluated at the jet sections where the stresses reach their maximum values. The maximum extensional stress is reached at $z\simeq -36$ $\mu$m. The axial coordinates corresponding to the maximum values of $\tau_s$ are shown in the figure. The numerical simulation was conducted for water and different liquid flow rates, as indicated in the figure.}
\label{stress2}
\end{figure}

The amount of processed material per unit time is proportional to the liquid flow rate. However, the efficiency of the exfoliation decreases as $Q_l$ increases. Figure \ref{stress2} shows the dependence of the viscous stresses on the liquid flow rate $Q_l$. The extensional stress $\tau_e$ is practically uniform across the (thinner) jet produced with the flow rate $Q_l=1.5$ ml/h selected in the experiments. Conversely, high values of $\tau_e$ are reached only near the interface when the flow rate increases. The maximum values are reached approximately at the same jet section in front of the discharge orifice ($z\sim -36$ $\mu$m). The shear viscous stress $\tau_s$ increases roughly linearly across the jet section for $Q_l=1.5$ ml/h, indicating the parabolic character of the velocity profile in this case. However, $\tau_s$ takes low values in the jet core for large liquid flow rates. 

As the liquid flow rate increases, the shear stress distribution becomes increasingly localized, and a larger fraction of the jet core remains weakly sheared. As a result, a significant portion of the liquid does not experience shear stresses high enough to promote exfoliation. This behavior can be rationalized considering flow-induced particle migration \citep{Poustka_2025}. Suspended particles are known to migrate away from regions of high shear toward low-shear zones, leading to accumulation near the channel centerline, where velocity profiles approach a plug-flow condition. As a consequence, at high flow rates, a significant fraction of particles is transported within the low-shear core of the jet, effectively bypassing the regions where shear stresses are sufficient to induce exfoliation. Overall, the results presented in Fig.~\ref{stress2} suggest that reducing the injected liquid flow rate may lead to an increase in the exfoliation efficiency because a larger fraction of the flow is exposed to significant extensional and shear stresses.

\begin{figure}
\begin{center}
\resizebox{0.5\textwidth}{!}{\includegraphics{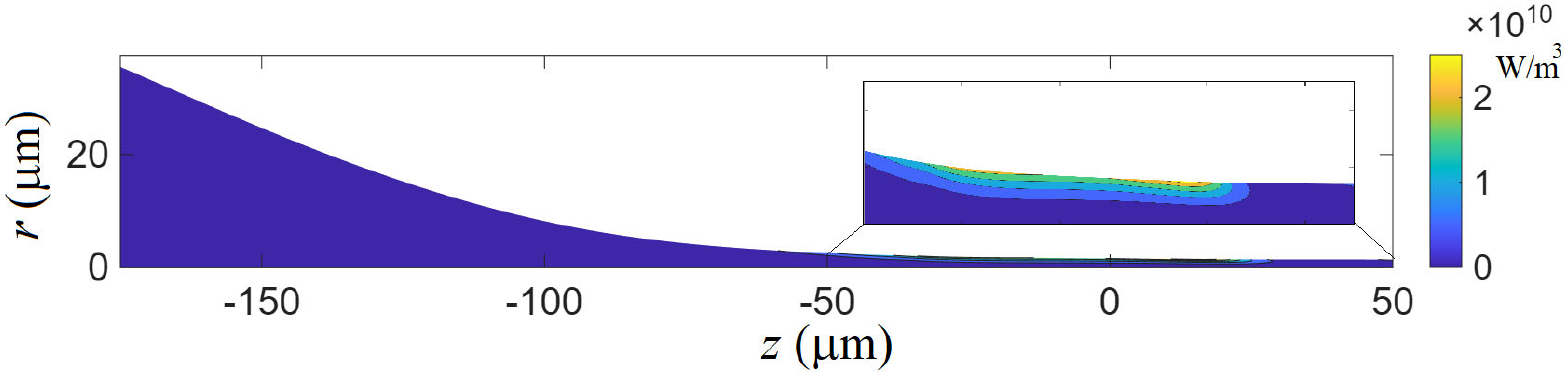}}
\end{center}
\vspace{0cm}
\caption{Viscous power density $\boldsymbol\tau:\nabla{\bf v}$ calculated from the numerical simulation for water and $Q_l=1.5$ ml/h.}
\label{stress02}
\end{figure}

\begin{figure}
\begin{center}
\resizebox{0.485\textwidth}{!}{\includegraphics{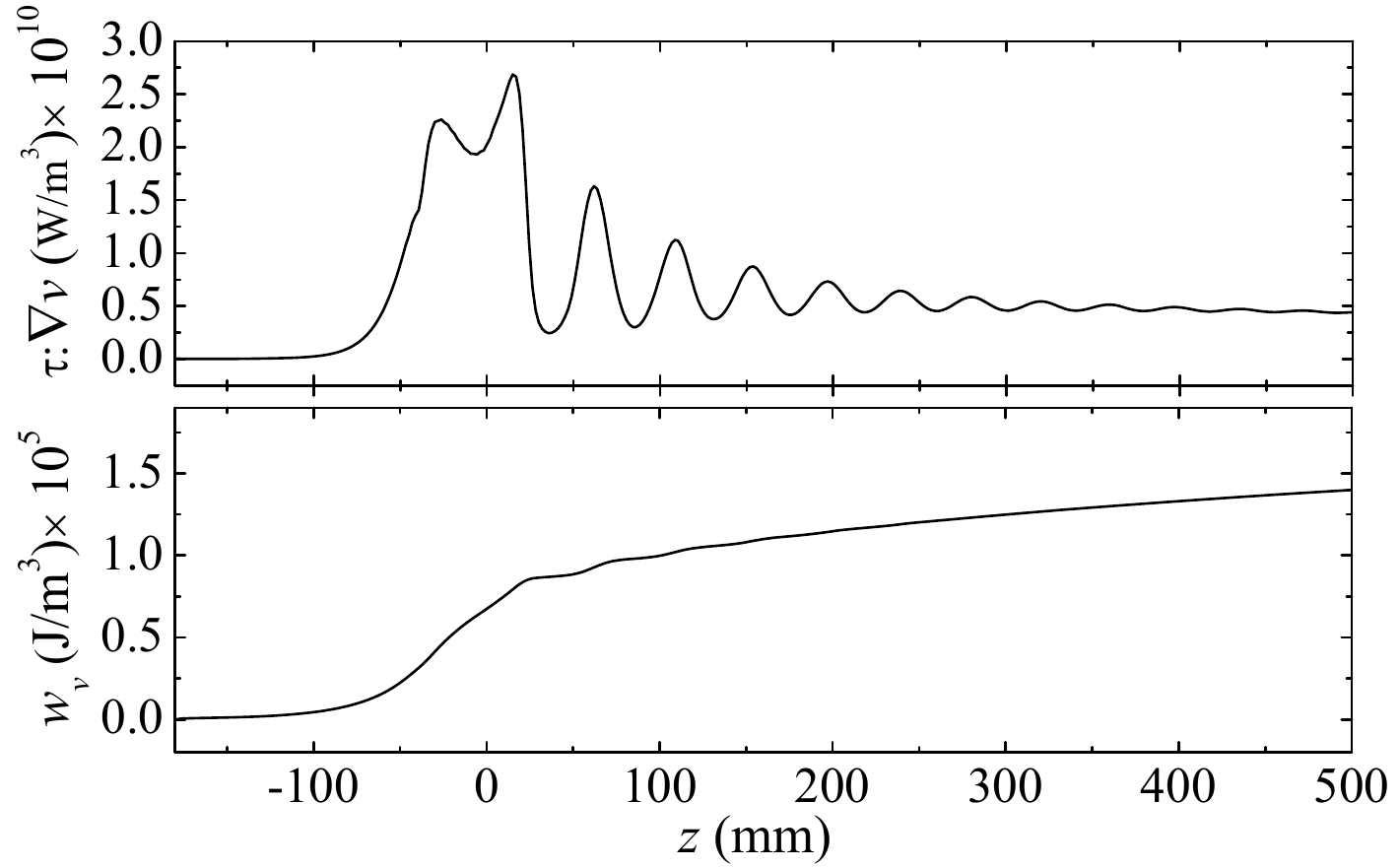}}
\end{center}
\vspace{0cm}
\caption{(a) Viscous power density $\boldsymbol\tau:\nabla{\bf v}$ along the interface. (b) Work per unit volume $w_v=\int_{0}^{s'} (\boldsymbol\tau:\nabla{\bf v})\, v^{-1} ds$ accumulated along the interface between $z=0$ and $z(s')$. Here, $s$ is the arclength of the interface. The numerical simulation was conducted for water and $Q_l=1.5$ ml/h.}
\label{stress12}
\end{figure}

The viscous power density $\boldsymbol\tau:\nabla{\bf v}$ sharply increases in the meniscus-jet transition, reaching values of the order of $10^{10}$ W/m$^3$ (Fig.\ \ref{stress02}). The viscous energy per unit volume of a fluid particle traveling along the interface sharply increases in that region (Fig.\ \ref{stress12}). The simulation result $w_v\sim 10^5$ J/m$^3$ is consistent with our scaling analysis. \revFJGR{Calculating the work done by the fluid stresses on the particle requires a two-phase simulation, which is beyond the scope of this paper.}  

\subsection{High-Yield Monolayer Exfoliation and Graphene \revFJGR{Nanofragments} Formation}\label{IIIC}

The analysis of images acquired with the high-speed video camera showed no disruption in liquid ejection. This suggests that the GNPs were smaller than the jet diameter, which was a few micrometers.

The morphology and thickness of the exfoliated nanosheets were characterized using High-Resolution Transmission Electron Microscopy (TEM) and Atomic Force Microscopy (AFM) in PeakForce Tapping mode. Full technical specifications, including cantilever calibration via the thermal K tune method and imaging setpoints, are detailed in Section II of the Supplementary Material.

\begin{figure}
\begin{center}
\resizebox{0.49\textwidth}{!}{\includegraphics{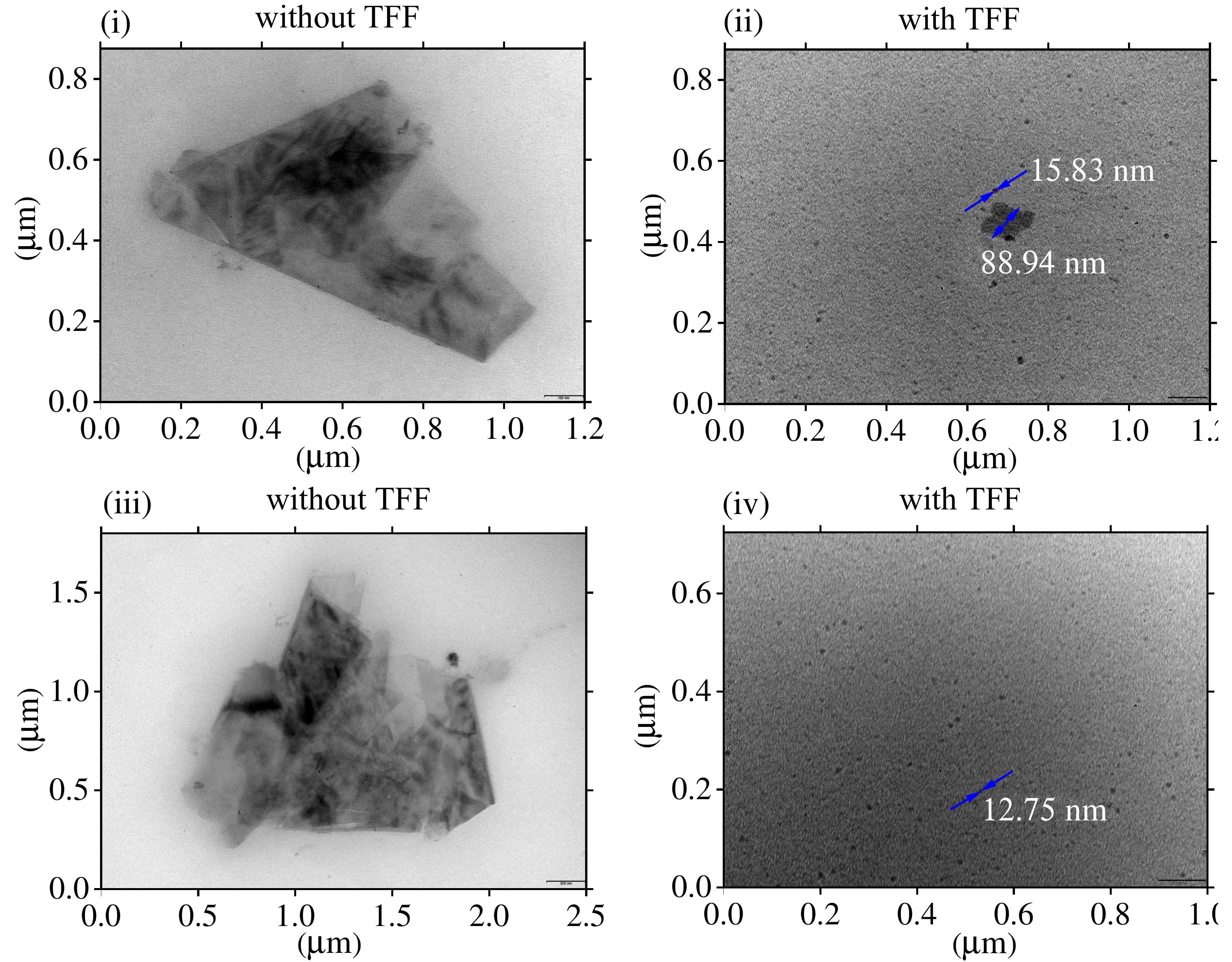}}
\end{center}
\vspace{0cm}
\caption{TEM images of graphene samples exfoliated from GNP using: (i) bath ultrasonication in isopropanol, (ii) bath ultrasonication followed by TFF in isopropanol, (iii) bath ultrasonication in pure water, and (iv) bath ultrasonication followed by TFF in pure water. Each panel shows a representative region of the deposited material after solvent evaporation. TFF induces a strong atomization of the starting platelets, yielding a clear transition from large multilayer flakes to a dense population of GQDs with typical lateral sizes of 10–15 nm.}
\label{TEM}
\end{figure}

Structural characterization via TEM (Fig.\ \ref{TEM}) reveals a \revFJGR{significant} morphological transformation of the starting graphite platelets upon TFF processing. While ultrasonication yielded only large, multilayer flakes (Fig.\ \ref{TEM}i, iii), the integration of TFF induces an efficient fragmentation of the precursor, producing a dense population of Graphene \revFJGR{nanofragments} with typical lateral sizes of 10–15 nm. Statistical analysis of the TEM micrographs confirms this consistency: in isopropanol, we detected a mean lateral size of $16.2\pm 3.1$ nm ($N=128$, Fig.\ \ref{TEM}ii), while in pure water, the mean size was $14.4\pm 2.5$ nm ($N=217$, Fig.\ \ref{TEM}iv). 

\begin{figure}
\begin{center}
\resizebox{0.49\textwidth}{!}{\includegraphics{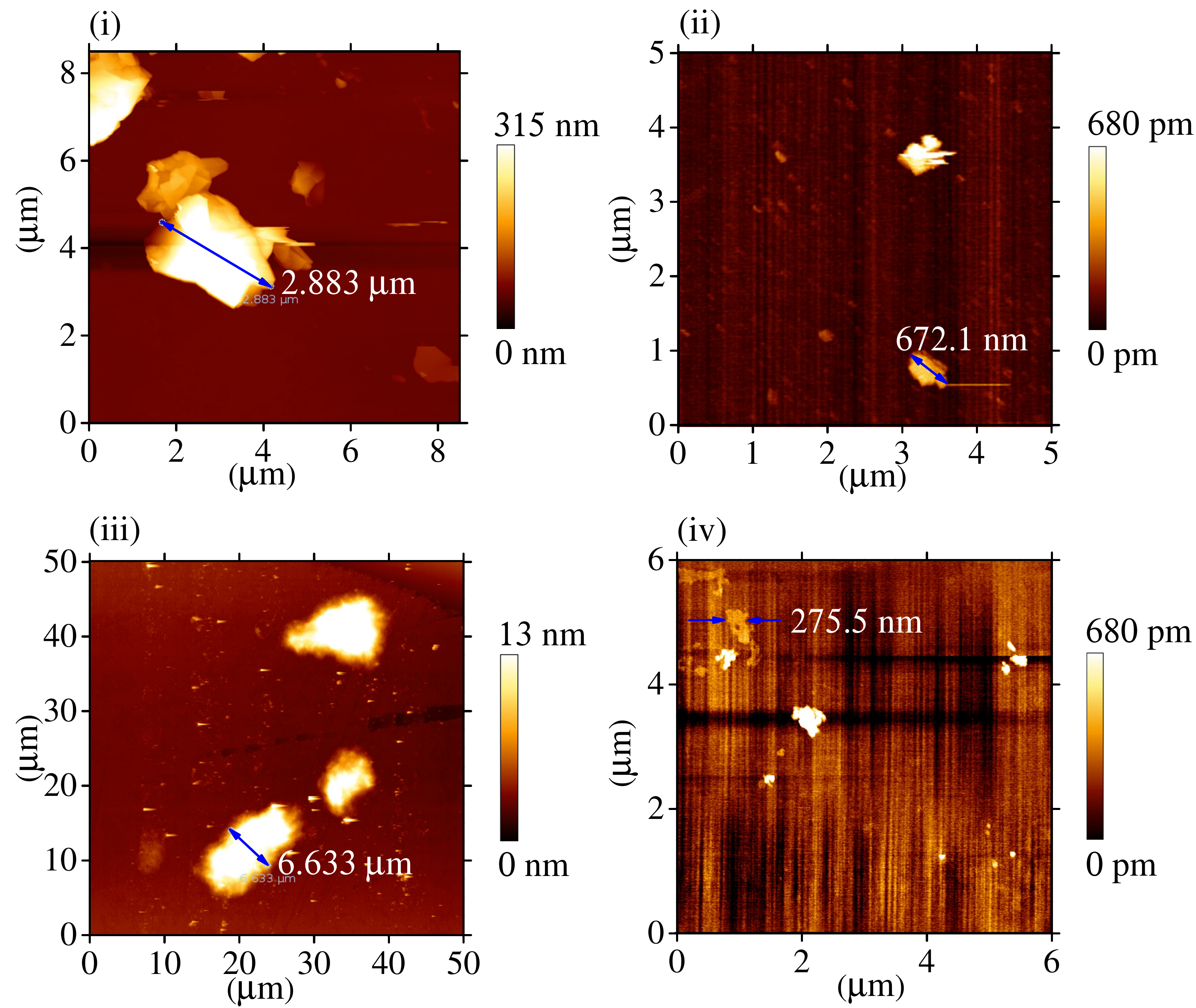}}
\end{center}
\vspace{0cm}
\caption{AFM images of graphene samples exfoliated from GNP using: (i) bath ultrasonication in isopropanol, (ii) bath ultrasonication followed by TFF in isopropanol, (iii) bath ultrasonication in pure water, and (iv) bath ultrasonication followed by TFF in pure water. Each image corresponds to a representative region of the deposited sample after solvent evaporation. The color scale represents the measured thickness (height). Particles with thicknesses exceeding the maximum scale value appear as white dots in the image. TFF produces monolayer graphene flakes (lateral size $\sim$ 400–500 nm) and monolayer GQDs (lateral size $\sim$10–15 nm) in both solvents in a \revFJGR{single TFF pass}.}
\label{AFM}
\end{figure}

Complementary AFM analysis (Fig.\ \ref{AFM}) further confirms the high selectivity of the TFF process, identifying both monolayer graphene flakes with lateral sizes in the range $400–500$ nm and monolayer GQDs with lateral dimensions in the interval 10–15 nm. To quantify exfoliation efficiency, we applied a height threshold of 680 pm to the AFM micrographs, corresponding to the theoretical bilayer thickness. Remarkably, 99.6\% of the particles detected in isopropanol ($N=276$) and 92.9\% in water ($N=238$) \revFJGR{exhibited apparent heights below this threshold, consistent with monolayer graphene. This high monolayer fraction}
, achieved in a single pass without the aid of surfactants or centrifugation, underscores the deterministic nature of the TFF platform.

\begin{figure}
\begin{center}
\resizebox{0.35\textwidth}{!}{\includegraphics{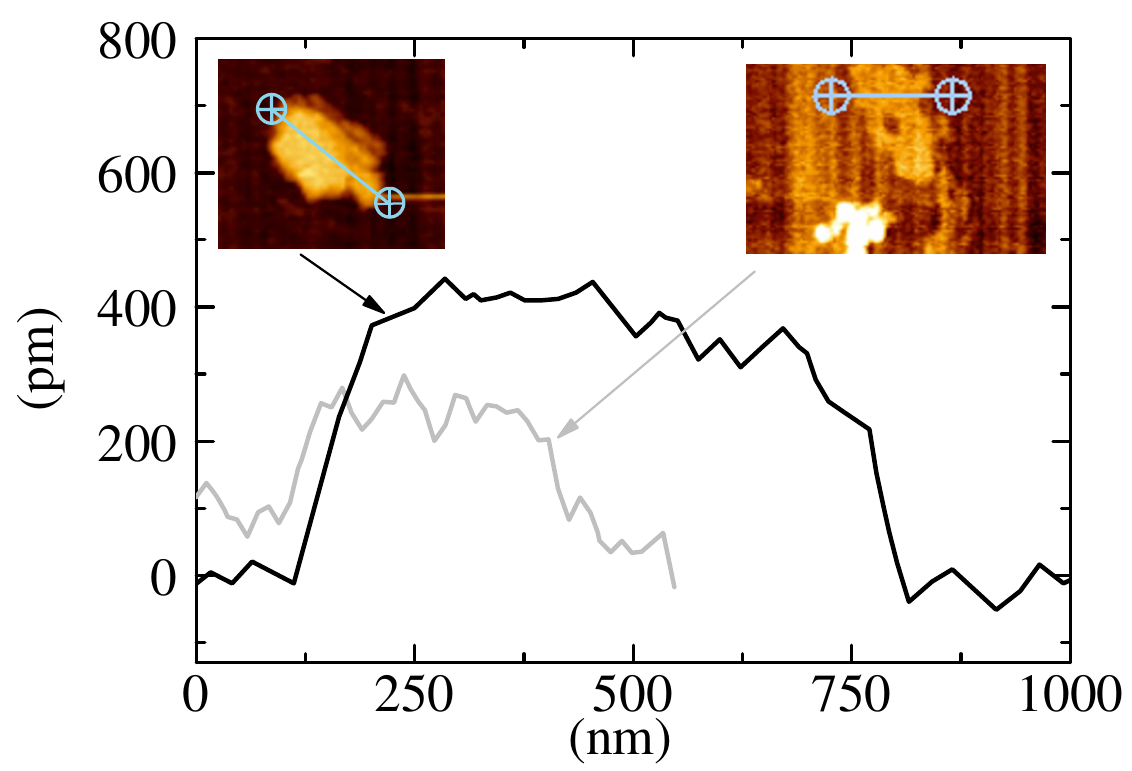}}
\end{center}
\vspace{0cm}
\caption{AFM profilometry of monolayer graphene flakes. Black and grey lines correspond to profiles of the monolayer graphene flakes indicated in Figs.\ \ref{AFM}-(ii) and (iv), respectively.}
\label{Perfilometry}
\end{figure}

The structural integrity and monolayer nature of the exfoliated material are further validated by high-resolution profilometry (Fig.\ \ref{Perfilometry}). The measured step heights for representative flakes in both isopropanol and water consistently fall within the 300–400 pm range (Figs.\ \ref{AFM}ii, iv), in excellent agreement with the theoretical thickness of monolayer graphene.


\section{Conclusions and Future Works}

We have demonstrated that TFF establishes a wall-free, deterministic paradigm for the high-throughput production of 2D and 0D nanomaterials. \revFJGR{Although their lateral dimensions fall within the range commonly associated with graphene quantum dots, optical evidence of quantum confinement will be addressed in future work.} By harnessing extreme extensional and shear stresses at a transonic meniscus, TFF achieves a \revFJGR{high} monolayer yield ($>90\%$) in a single-pass process, simultaneously generating graphene flakes and \revFJGR{nanofragments} from the same precursor. Scaling analysis and numerical simulations have allowed us to rationalize our experimental results.

Unlike conventional liquid-phase exfoliation, the TFF platform operates in green solvents, including pure water, without \revFJGR{surfactant addition, oxidative treatment, or centrifugation fractionation. }
The absence of solid-wall interactions \revFJGR{is expected to reduce abrasion-related damage and, consequently, preserve} the structural integrity of the nanosheets, providing an application-ready output. \revFJGR{However, a full assessment of defect density would require a further spectroscopic characterization}. Given its compact, modular design, TFF offers a straightforward path to parallelization and industrial-scale nanomanufacturing. 


\section{Materials and Methods}
\label{mat}

\subsection{Experimental method}
\label{mat:expmet}

Here, we briefly describe the process of ejector fabrication, sample preparation, and sample characterization. Further details are provided in the Supplemental Material. The TFF ejector was fabricated via Dip-in Laser Lithography (DiLL) using a Nanoscribe Photonic Professional GT2 system with IP-S photoresist, following the procedure described by Rubio et al.~\citep{RRCHGM21}. 

Suspensions of graphene nanoplatelets (GNPs) were supplied by Graphenest SA (Aveiro, Portugal) \citep{MARFFFS20}. The method for producing the GNPs comprises dispersing the particles in a solvent, subjecting the resulting dispersion to a cavitation force sufficient to generate cavitation bubbles, and
subjecting the dispersion to high-shear agitation. As a result, the GNPs with lateral sizes in the range $3.2-8.5$ $\mu$m were moderately pre-expanded with an interlayer distance around $500-700$ pm and adhesion energy $\gamma_{\textin{eff}}\sim 10^{-2}$ J m$^{-2}$.

GNPs were prepared in both water (density $\rho=998$ kg$\cdot$m$^{-3}$, viscosity $\mu=$1 mPa$\cdot$s, and surface tension $\sigma=72$ mN/m) and isopropanol ($\rho=785$ kg$\cdot$m$^{-3}$, $\mu=2.0$ mPa$\cdot$s, and $\sigma=21.7$ mN/m). Ultrapure water was obtained from a {\sc Direct-Q\textregistered 3} purification system, and 99.9\% isopropanol was supplied by Sigma Aldrich\textregistered. 

GNP dispersions (230~ppm) were processed in an ultrasonic bath (40~kHz and 120~W) for three hours, ensuring that the bath temperature did not exceed \SI{50}{\celsius} \citep{RCAG23,RG24}. To prevent nozzle clogging, both dispersions were passed through a 10 $\mu$m Acrodisc\texttrademark PSF syringe filter. Post-filtration, small volumes were drop-cast onto freshly cleaved mica substrates for AFM and onto Formvar/carbon-coated copper grids for TEM. The remaining filtered dispersions were then subjected to TFF. The resulting exfoliated samples were again cast for TEM and AFM analysis to quantify the transition from bulk platelets to monolayers and GQDs. 

\subsection{Numerical method}
\label{sec:MatMetNum}
We integrate the conservation equations for mass, momentum, and energy for the two phases:
\begin{equation}
\frac{\partial \rho}{\partial t} + \nabla \cdot (\rho {\bf v})=0, \quad 
\rho\frac{D{\bf v}}{Dt}  = -\nabla p+\nabla\cdot\boldsymbol \tau,
\end{equation}
\begin{equation}
\label{e}
\rho\frac{D(c_vT)}{Dt}=-p\nabla\cdot {\bf v}+\boldsymbol\tau:\nabla{\bf v} - \nabla\cdot{\mathbf q}.
\end{equation}
Here, $\rho({\bf r},t)$, ${\bf v}({\bf r},t)=v_r({\bf r},t) {\bf e_r}+v_z({\bf r},t) {\bf e_z}$, $p({\bf r},t)$, and $T({\bf r},t)$ are the axisymmetric density, velocity, pressure, and temperature fields in each phase, respectively, $c_v$ is the specific heat capacity at constant volume, and $D/Dt$ is the material derivative. These equations are completed with the constitutive relationships for the viscous stress tensor $\boldsymbol \tau$ and the heat flux vector ${\bf q}$: 
\begin{equation}
\boldsymbol \tau =  \mu \left(\nabla{\bf v}+ (\nabla{\bf v} )^\mathrm{T}\right)+\lambda (\nabla\cdot{\bf v}) \mathbf I, \quad {\bf q}=-\kappa \nabla T,
\end{equation}
where $\mu$, $\lambda$, and $\kappa$ represent the shear viscosity, dilatational coefficient of viscosity, and thermal conductivity of each phase, respectively, and ${\bf I}$ is the identity matrix. In addition, the equation of state $p=\rho R_g T$ is considered in the gas phase, where $R_g=c_p-c_v$ is the gas constant and $c_p$ is the specific heat capacity at constant pressure. We assume that the liquid is incompressible, i.e. $\nabla \cdot {\bf v}=0$ in the liquid phase. 

We impose the continuity of velocity, temperature, stress, and heat flux at the interface:
\begin{equation}
\label{stred}
||{\bf v}||=0,\quad ||T||=0,\quad \mathbf{n}\cdot ||{\boldsymbol \tau}||-||p||\mathbf{n}=\sigma (\nabla\cdot {\bf n}) {\bf n},
\end{equation}
\begin{equation}
||\kappa\, \partial T/\partial n||=0,
\end{equation}
where $||A||$ denotes the difference between the values taken by the quantity $A$ on the two sides of the interface, ${\bf n}$ is the unit outward normal vector, and $\sigma$ is the surface tension. Also, we consider the compatibility condition
\begin{equation}
\frac{\partial F}{\partial t}-v_r+v_z\frac{\partial F}{\partial z}= 0\; ,
\end{equation}
where $F(z,t)$ is the distance of an interface element from the axis of symmetry $z$. 

We prescribe parabolic and uniform axial velocity profiles at the liquid and gas inlets, respectively. The stagnation pressure $p_0$ and temperature $T_0$ are fixed at the gas inlet, while the flow rate $Q_l$ and temperature $T_0$ are set at the inlet of the liquid feeding capillary. The zero-gradient (outflow) boundary condition is imposed at the gas and liquid outlets for all the variables except for the pressure, whose value $p_e$ is fixed at that section. The no-slip ${\bf v}={\bf 0}$ and no-temperature jump $T=T_0$ boundary conditions are imposed on the solid surfaces. We verified that the results are practically the same if the condition $T=T_0$ is replaced by the adiabatic wall boundary condition $\partial T/\partial n=0$ on the solid surfaces. We did not consider the latter because it hinders the convergence of the solution. To complete the set of boundary conditions, we assume $\partial p/\partial n=0$ on both the solid surfaces and the free surface. 

The flow is calculated using a variant of the boundary-fitted method described by Montanero and Herrada \citep{HM16a,HERRADA25}. A quasi-elliptic transformation \citep{DT03} is applied to generate the grid, which allows us to deal with the sharp reduction of the free surface radius in the meniscus tip. The equations are discretized in the transformed radial direction $\eta$ using $n^{\ell}_{\eta}=n^{g1}_{\eta}=15$ and $n^{g2}_{\eta}=21$ Chebyshev collocation points \citep{K89} in the liquid and gas domains, respectively. The transformed axial direction $\xi$ is discretized using fourth-order finite differences with $n^{\ell}_{\xi}=n^{g1}_{\xi}=1345$ and $n^{g2}_{\xi}=673$ equally spaced points. The grid points accumulate near the free surface, enabling accurate integration of the gaseous viscous boundary layer. 

\section*{Acknowledgements}
Financial support from the Spanish Ministry of Science and Education (Grant no. PID2019-108278RB-C32 / AEI / 10.13039/501100011033) and the Gobierno de Extremadura (Grant no. GR18175) is gratefully acknowledged. This work was also supported by the Fundação para a Ciência e a Tecnologia (FCT) and the Ministério da Educação, Ciência e Inovação (MECI) through projects UID/00532/2025, UID/PRR/00532/2025, LA/P/0045/2020, and 2020.03203.CEECIND. The authors would like to thank Dr. Manuela Brás (Advanced Biomaterials and Biointerfaces Characterization scientific platform - ABC), Dr. Sofia Pacheco, and Dr. Rui Fernandes at I3S, University of Porto, Portugal, for their expert assistance with the AFM and TEM characterization. Finally, the authors are grateful to Graphenest SA for graciously supplying the graphene nanoparticles used in this study.

\section*{Data Availability}
The data that support the findings of this study are available from the corresponding author upon reasonable request.

Supplementary Information is available for this paper.\\

\newpage


\end{document}